%% file: main.tex
\pdfoutput=1
\documentclass[letterpaper]{article}
\usepackage{aaai20}
\usepackage{times}
\usepackage{helvet}
\usepackage{courier}
\usepackage{latexsym}
\usepackage{enumerate}
\usepackage{enumitem}
\usepackage{url}

\usepackage{booktabs} % For formal tables
\usepackage{graphicx}
\graphicspath{ {./plots/} }
\usepackage{subfig}
\usepackage{mathtools}
\usepackage{array}

\usepackage{soul}
\usepackage{xcolor}
\usepackage{comment}

\newcommand{\mysection}[1]{
\section{#1}
\vspace{-0.1cm}
}

\begin{document}

\author{Moniba Keymanesh\textsuperscript{1}, 
Saket Gurukar\textsuperscript{1}, 
Bethany Boettner\textsuperscript{1}, \AND 
Christopher Browning\textsuperscript{1},
Catherine Calder\textsuperscript{2},\and
Srinivasan Parthasarathy\textsuperscript{1} \\ \\
\textsuperscript{1}{The Ohio State University},
\textsuperscript{2}{University of Texas at Austin}\\
\{keymanesh.1, gurukar.1, boettner.6, browning.90, parthasarathy.2\}@osu.edu,\\
calder@austin.utexas.edu}

\title{Twitter Watch: Leveraging Social Media to Monitor and Predict Collective-Efficacy of Neighborhoods}
%\titlenote{The name of the paper, "Twitter Watch" is inspired by the Neighborhood Watch programs that are organized group of civilians devoted to crime and vandalism prevention within a neighborhood}
\maketitle

\begin{abstract}
\input{abstract.tex}

\end{abstract}

%\keywords{}

%%%%%%%%%%%%%%%%%%%%%%%%%%%%%%%%%%%%%%%%%%%%%%%%%%%%%%%%%%%%%%%%%%%%%%%%%%%%%%%%

\mysection{INTRODUCTION}
\label{s:1}
\input{Introduction.tex}

\mysection{Background and Related Work}
\label{s:Background and Related Work}
\input{relatedwork.tex}

\mysection{Data Collection}
\label{s:DataCollection}
\input{data.tex}
%\section{Forecasting Collective efficacy}
\mysection{Methodology}
\label{s:Forecasting Collective efficacy}
\input{method.tex}
\mysection{EXPERIMENTS}
\label{s:EXPERIMENTS}
\input{expriments.tex}

\mysection{Results}
\label{s:Results}
\input{results.tex}
%\section{Discussion and Future work}
%\label{s:Discussion and Future work}
%\input{discussion.tex}
\mysection{CONCLUSION}
\label{s:CONCLUSION}
\input{conclusion.tex}

\section{Acknowledgements}
\label{s:Acknowledgements}
\input{ack.tex}
\bibliographystyle{aaai}
\bibliography{mybib} 
\clearpage

\appendix
\mysection{Supplemental Material}
\label{s:suplementary}
\input{Suplementary_material.tex}

\end{document}

%% file: abstract.tex
%The occurrence of criminal violence is uneven across urban communities. 
Sociologists associate the spatial variation of crime within an urban setting, with the concept of collective efficacy. The collective efficacy of a neighborhood is defined as social cohesion among neighbors combined with their willingness to intervene on behalf of the common good. Sociologists measure collective efficacy by conducting survey studies designed to measure individuals' perception of their community. In this work, we employ the curated data from a survey study (ground truth) and examine the effectiveness of substituting costly survey questionnaires with proxies derived from social media. We enrich a corpus of tweets mentioning a local venue with several linguistic and topological features. We then propose a pairwise learning to rank model with the goal of identifying a ranking of neighborhoods that is similar to the ranking obtained from the ground truth collective efficacy values. In our experiments, we find that our generated ranking of neighborhoods achieves 0.77 Kendall tau-x ranking agreement with the ground truth ranking. Overall, our results are up to 37\% better than traditional baselines.

%% file: Introduction.tex
Understanding occurrence of crime and disorder in cities is important for public health, policy, and governance. However, occurrence of criminal violence is uneven across the neighborhoods~\cite{chainey2013gis,weisburd2009units}.
%Moreover, as the neighborhood changes with time the frequency of disorder in these neighborhood changes~\cite{kirk2010neighborhood}. 
Sociologists and policy-makers associate the spatial variation of disorder to the organizational characteristics of the neighborhoods~\cite{morenoff2001neighborhood,sampson1997neighborhoods,kornhauser1978social,sampson1989community,browning2016neighborhood}. An important measure of such disorder is collective efficacy \cite{sampson1997neighborhoods}. \textit{Collective efficacy} is defined as ``social cohesion and trust among neighbors combined with the joint willingness to intervene on behalf of the common good" ~\cite{sampson1997neighborhoods}. Collective efficacy is increasingly used by local governments to prioritize resources to both monitor and reduce disorder through targeted policies and neighborhood gentrification strategies. It is also used to measure the impact of said policies and strategies over time~\cite{hipp2016collective,bandura1997editorial}.
%As the neighborhood changes with time, collective efficacy in a neighborhood changes through an \textit{updating} process~\cite{hipp2016collective}. % SAKET: THIS CAN GO IN RELATED WORK: ``Updating" in collective efficacy context refers to the process in which neighborhood members reassess their sense of efficacy and perception of social cohesion based on the new information that has been acquired~\cite{bandura1997editorial}.

The computation of neighborhoods'\footnote{In this paper we use the terms ``neighborhood" and ``block group" interchangeably. Block group refers to a census block group which is a smallest geographical unit for which the United States census bureau publishes sample data.} collective efficacy traditionally requires conducting expensive surveys; usually requiring funding on the order of hundreds of thousands of dollars~\cite{couper2017new}. 
Changes to collective efficacy over time~\cite{hipp2016collective}, due to policy shifts (e.g. through neighborhood gentrification efforts) require additional surveys, further exacerbating this cost.

%With the goal of mitigating this cost, we ask the following question: ``Can we leverage social media data targeting neighborhoods of interest to estimate the collective efficacy of the neighborhoods?".

%Many studies have shown the effectiveness of leveraging social media data in predicting depression \cite{de2013predicting}, estimating heart disease mortality \cite{eichstaedt2015psychological}. 

Sociologists and government agencies typically use collective efficacy to ``order" neighborhoods with respect to neighborhood safety perception, social cohesion among residents, and their  willingness to intervene on behalf of the common good.    Neighborhoods with high collective efficacy tend to be safer while lower collective efficacy values correspond to relatively less safe neighborhoods~\cite{sampson1997neighborhoods}. 
Essentially one may model this as a ranking problem.  Concretely the key question we seek to answer in this paper is: ``Given the social media data about neighborhoods, can we rank the neighborhoods such that the ranked list is close to the ranked list of neighborhoods ordered by collective efficacy -- thereby saving on the cost of expensive surveys?". 

Our approach, a first of its kind study at a city-scale, seeks to characterize  neighborhood collective efficacy by levering spatially conditioned linguistic features extracted from social media. These features are related to the type of urban activity, language use, visible signs of crime and anti-social behavior reported on such media, familiarity of residents with one another, and public mood of the neighborhood. We lever additional sociological, and spatial features and develop a simple pairwise learning to rank model based on these features. We empirically show the effectiveness of our model on a real world city-scale dataset, with ground truth values of collective efficacy computed from a traditional survey-based study~(details in section~\ref{ahdcdata}). Additionally, we conduct a comprehensive analysis of the predictive power of specific features in the learning to rank task to better understand the relative importance of individual features. In terms of broader impacts such ideas can be used as a cost-effective early warning mechanism to monitor the transformations of the neighborhoods and prioritize the resources.

%% file: relatedwork.tex
% Collective efficacy of a neighborhood refers to the ability of the members of neighborhood to maintain social control by controlling the behaviour of individuals and groups in the neighborhood \cite{sampson1997neighborhoods}. As mentioned in the previous section, collective efficacy of a neighborhood changes with time through an ``updating process". ``Updating" in collective efficacy context refers to the process in which neighborhood members reassess their sense of efficacy and perception of social cohesion based on the new information that has been acquired~\cite{bandura1997editorial}.

Twitter has been used by researchers to make sense of human behavior. The behavior of users on this platform has been used in particular to assist prediction of criminal violence~\cite{wang2012automatic,gerber2014predicting,wang2012spatio,wang2015using,aghababaei2016mining,williams2017crime,bendler2014investigating}. The link between the prediction of social unrest and the user's online activity on Twitter has been studied by~\cite{compton2013detecting}. Moreover, Twitter has been employed to study the online behavior of gang members~\cite{patton2015gang} and to measure the population at risk, considering violent crime~\cite{malleson2015impact}. Several studies have used Twitter to study the trust relations~\cite{vedula2017predicting} among online users. 
%Twitter has shown promising results for trust prediction in the context of emergent real-world crisis scenario~\cite{vedula2017predicting}.
Researchers have also leveraged Twitter data for studying social disorganization by evaluating entropy of individuals' opinion about soccer teams~\cite{pacheco2017using}. Although the concepts of trust, crime, and social disorganization are related to collective efficacy, to the best of our knowledge estimating individuals' perception of their social climate and expectation of intervention using social media data has not been addressed till now.

%% file: data.tex
\begin{figure}
 \includegraphics[scale=1]{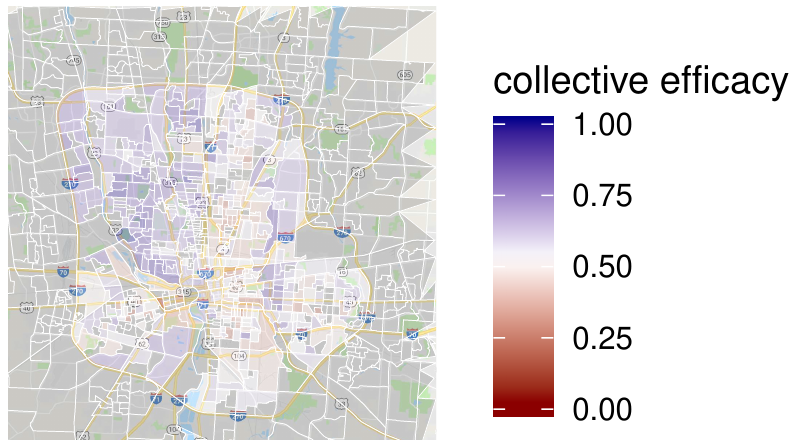}
\centering
\caption{Collective efficacy map of Columbus, OH}
\label{fig:map}
\end{figure}

\label{ahdcdata}
\textbf{AHDC study:} The adolescent health and development in context~(AHDC) study is a longitudinal data collection effort in a representative and diverse urban setting that focuses on the contribution of social and spatial environments to the health and developmental outcomes of urban youth. The study area is a contiguous space in Columbus Ohio. In the first wave of the study 1403 Columbus residents participated in the study.
Participants were asked a series of questions about their neighborhood and routine activity locations. Questions specifically focused on informal social control items measuring the participant's perception of the social climate in the area at and around each location and in the neighborhood. Participants reported agreement with the following questions: 1- whether people on the streets can be trusted? 2- whether people are watching what is happening on the street?, and 3- whether people would come to the defense of others being threatened? Responses ranged from 1 (``strongly disagree") to 5 (``strongly agree"). This step resulted in roughly 9000 location reports (4031 unique locations) nested within 567 block groups. In order to achieve the collective efficacy value of each neighborhood,  we aggregated individual responses to the three social control items at the report level. Then, we aggregated report-level results for each block group. Finally, we normalized the scores in range of 0 to 1 and use it as the ground truth for our study. This methodology is aligned with the traditional measurement approach employed to compute collective efficacy at neighborhood level~\cite{bandura1997self,paskevich1999relationship,sampson1999beyond}.
Note that while we adopt a similar ground truth model~\cite{sampson1999beyond}, we lever survey reports from individuals who {\it both} reside within {\it and frequently visit} a particular neighborhood (the original model focused just on residents). Concomitantly, the social media postings included in our study includes postings from both individuals that reside within and frequently visit a particular neighborhood. 
In order to increase the reliability of the aggregation we only include the neighborhoods having at least 5 reports. Figure~\ref{fig:map} shows a collective efficacy map of Columbus (ground truth). 

\textbf{Twitter Data:} 
Matching the offline survey instrument of the AHDC study we collected a significant corpus of Twitter feeds from Columbus area and its suburbs. Our goal was to capture the informal language of local citizenry focused on local venues and localities.
We chose Twitter because of its national appeal\footnote{http://www.pewinternet.org/2018/03/01/social-media-use-in-2018/}  and easy availability through the API. For the purpose of our study, collectively, more than 50 million publicly available tweets were collected from the accounts of 54k Twitter users who identified their location as Columbus. These users were identified through Snowball sampling~\cite{goodman1961snowball}. Details of our data collection process can be found in Appendix~\ref{datacollection-suplementary}.
%63 Twitter accounts that mostly posted news and information about MUC city were identified and used as the seed users. Many local residents follow such accounts to stay informed about the local events~\cite{kwak2010twitter}. The seed set included the twitter account of several organizations including major universities, recreational centers, medical centers, newspapers, local bloggers, local reporters, police, libraries, restaurants as well as the local sports teams. Using Twitter's streaming API, the followers of the seed accounts were collected. Following this step, we explored user's profiles and identified 54K public profiles that marked their locations as MUC city or one of the suburban areas included in the PSC study. The PSC study area included several populous suburbs.  
%\textit{Whitehall},\textit{ Bexley}, \textit{Upper Arlington}, \textit{Grandview Heigths}, \textit{Worthington}, \textit{Hilliard}, and \textit{Dublin}.  
%Collectively, 50 million publicly available tweets were collected from these accounts. In another wave of data collection, we collected geo-tagged tweets for a period of May-August of 2018. This resulted to additional 2.8 million tweets. 

\subsection{Associating Tweets to Neighborhoods}
\label{s:data collection}

Following our data collection, we excluded the tweets that did not contain a mention of locations within our study area. %In order to identify the tweets that mention the locations in our study,
For doing so, we used a state-of-the-art publicly available location name extractor LNEx~\cite{al2017location}. LNEx extracts location entries from tweets, handles abbreviations, tackles appellation formation and metonomy pose disambiguation problems given gazetteer and region information. 
Open Street Map gazetteer was used and region was set to Columbus.
%LNEx is designed to extract location entities from informal and unstructured social media data. 
%- to identify the tweets which were mentioning a location.
%Subsequently, Open Street Map gazetteer was used for geocoding and associating the venues mentioned in the tweets to a neighborhood in the MUC area.
However, there are cases in which ambiguous locations were reported by LNEx. In our study, we exclude tweets containing ambiguous location entities. For more details on the ambiguous cases and pruning steps see Appendix ~\ref{datacollection-suplementary}. 
% The location ambiguities were observed in a following cases: 
% \begin{itemize}[noitemsep,topsep=0pt]
%     \item A location entity may have several matches in the gazetteer. For example, \textit{Holiday Inn} and \textit{Gamestop}. 
%     \item A location entity having a single gazetteer entry can potentially refer to a huge area. For example, gazetteer entry of Trans-Siberian Highway in Russia spans from St. Petersberg to Vladivastok. Such entities cannot be mapped to a single neighborhood or block group. 
%     \item Location entities extracted by LNEx having a gazetteer entry but not referring to a location in the context. For example, \textit{American Girl}, \textit{Modern Male} etc. 
% \end{itemize}
%For example, \textit{Olentangy trail}, \textit{Ackerman Road}, and Franklin County. 
%III) LNEx extracted some non-location entities.
%For example, \textit{Basketball} and \textit{Drinking water}
%Such mentions were identified manually and excluded from the study. 
This pruning step resulted in 4846 unique locations that were spotted in 545k tweets and were mapped to 424 neighborhoods.\footnote{It is our intent to open source the tweet IDs as well as the ground truth collective efficacy values of neighborhoods in Columbus once this work is published.}

%% file: method.tex
In this section, we formalize our prediction task and the proposed ranking model. As mentioned earlier, the goal of our study is to rank the neighborhoods based on features extracted from tweets such that the ranked list is similar to the list of neighborhoods ordered by collective efficacy. Hence, we formulated our problem as pairwise learning to rank task~\cite{liu2009learning}.  

 \subsection{Definitions and Problem Formulation}
 \label{problem_formulation}
 First, we  define the terms related to ordinal ranking. \textit{Tied objects} refer to the set of two or more objects that are interchangeable in ranking with respect to the quality under consideration~\cite{kendall1945treatment}. %Following~\cite{cook1986axiomatic}, we define \textit{linear ordering} as one in which all the objects are ranked and no ties are permitted. Note that in a ranking in which ties are not permitted all pairs of objects are compared and form a totally ordered set while 
 %In a ranking in which ties are allowed, tied objects are not compared and thus, they form a\textit{ partially ordered set(poset)}. 
 The ranking in which ties are allowed is called a \textit{weak ordering}.
 %If the restriction regarding no ties is removed from the linear ordering to allow ties the resulting ranking is called \textit{weak ordering}.%Our goal is to automatically generate a ranking of neighborhoods using features extracted from tweets in a way that the level of social control of the neighborhood decreases throughout the list.
In our study, the neighborhoods with significantly small difference in their ground truth value of collective efficacy are considered \textit{tied}. As a result they are interchangeable in ranking. %However, neighborhoods having a significantly small difference in their ground truth value of collective efficacy are not to be compared against each other. In other words, they are interchangeable in ranking. 
%We account for this constraint by defining our ranking task as predicting a \textit{weak ordering} in which all the neighborhoods are ranked. %and unlimited ties can be defined. 
Ties are defined based on a threshold on the difference of collective efficacy values. Note that in this case, ties are intransitive by definition. Meaning that a tie relationship between neighborhoods $n_a$ and $n_b$; and $n_b$ and $n_c$ does not imply that neighborhoods $n_a$ and $n_c$ are tied. This constraint will be reflected in the way we define the ranking matrix and will be discussed in section~\ref{eval}. % plz do not change this to metric I mean the score matrix that we will define later

 Next, we formalize our ranking task and our proposed approach. Our data consists of \{$t_{1},t_{2},...,t_{c}$\} where $t_{i}$ is the set of tweets associated with neighborhood $n_i$. We denote the collective efficacy of neighborhood $n_i$ with $C(n_i)$. The goal of our framework is to automatically generate a permutation of neighborhoods $(f(n_1)f(n_2).....f(n_m) )$ where $f$ is the ordering function that maps each neighborhood to its position such that the mapped position of neighborhood is close to its true position based on  collective efficacy values.
 Formally the ranking task is defined as \{$f(n_i)~$\textless$~f(n_j) \mid \forall n_i, n_j$ if $C(n_i)~\leq~C(n_j)$\}.
 
 In order to generate a ranking of the neighborhoods, we first predict the local rank of all pairs of neighborhoods $n_i$ and $n_j$. In this scenario, there can be three cases for any pair; $n_i$ comes before $n_j$, $n_i$ comes after $n_j$, or $n_i$ and $n_j$ are interchangeable in the ranking. Thus, we formulate our local ranking task as a 3-class classification task. We then use the local rankings to generate the global ranking. Details of this process are discussed in Section~\ref{s:specific setup}. Next, we explain the features used in this study to characterize the neighborhoods. 

%  \begin{enumerate}[nosep, topsep=1pt]
%      \item $n_i$ comes before $n_j$ in the ranking.
%      \item $n_i$ comes after $n_j$ in the ranking.
%      \item $n_i$ and $n_j$ are interchangeable.
%  \end{enumerate}

 \subsection{Features}\label{s:features}
 
We characterize each neighborhood with features extracted from the tweets associated with the neighborhood. We compute two types of features: I) features that are computed for each neighborhood and II) features that are computed for a pair of neighborhoods. 
To generate the feature vector of a pair of neighborhoods, we first concatenate feature vector of each of the neighborhoods. Next, we add the pairwise features to the feature vector. These features are explained in detail in the following subsections.

 \subsubsection{\textbf{TF-IDF of crime related words:}} 
 ``Broken Windows"~\cite{wilson1982broken} is a well-known theory in criminology. The basic formulation of this theory is that visible signs of crime creates an urban environment that encourages further crime and disorder~\cite{skogan2015disorder,welsh2015reimagining}. Under the broken windows theory, %an ordered and well-maintained environment indicates that the area is monitored and criminal behavior is not tolerated. On the other hand, 
 a disordered environment, with signs of broken windows, graffiti, prostitutes, and excessive litter sends the signal that the area is not monitored and that criminal behavior has little risk of detection. Such a signal can potentially draw offenders from outside of the neighborhood. 
 On the basis of this theory, we used a lexicon of crime\footnote{This lexicon has been acquired from an open source repository https://github.com/sefabey/fear\_of\_crime\_paper} as a proxy for visible signs of crime and disorder in neighborhoods. This lexicon contains words that people often use while talking about crime and disorder.
%  Figure~\ref{fig:heatmap} shows the frequency of crime lexicon usage in two groups of neighborhoods. The top 20 neighborhoods in the plot have a high level of collective efficacy. While the bottom 20 neighborhoods have the lowest degree of collective efficacy in the area of our study. It can be seen that visible signs of crime are more frequent in neighborhoods with lower level of collective efficacy. This observation motivates using frequency of crime lexicon in tweets as a feature for characterizing the neighborhoods.
 TF-IDF captures the importance of a term in a document. With this in mind, we employed TF-IDF to capture the content surrounding the location entity in a tweet. For more details of preprocessing see the Appendix~\ref{tfidf-supplementary}.
%  We tokenized each tweet in our train set preserving the hashtags, handles, and emojis as separate words. We then removed the stopwords and lemmatized the tokens. Bigrams of the tweets were added to the token set. The top 100 crime-related terms that had the most frequency across the tweets were chosen as our vocabulary set. For the test set, we concatenated all the tweets in each neighborhood to get a single corpus per each neighborhood. We then, transformed each corpus to get the corresponding term-document vector.  

 \subsubsection{\textbf{Distribution of spatio-temporal urban activities using topic modeling:}}
 
  Casual, superficial interaction and the resulting public familiarity engender place-based trust among residents and ultimately the expectation of response to deviant behaviour~\cite{jane1961death}. Identifying the activities that individuals conduct in a city is a non-trivial step to understanding the ecological dynamics of a neighborhood such as the potential for street activity and public contact. Following the same methodology as in~\cite{fu2018identifying} we applied Latent Dirichlet Allocation~(LDA)~\cite{blei2003latent} to tweets associated with a given neighborhood to identify the main activity types in each neighborhood.
  %LDA treats words in a set of tweets as discrete signals and utilizes the word frequency distribution among tweets as statistical features. Each topic is characterized by a unique probability distribution of the vocabulary that is used. LDA assumes that a set of $T$ tweets involves $K$ latent topics. The output of this approach is the word distribution for each topic and topic distribution for a set of tweets. 
  The number of topics in a set of tweets is an important prior parameter in LDA model. To evaluate the topic model and determine the optimal number of topics, \textit{perplexity}~\cite{blei2003latent} is used by convention in language modeling. Perplexity is defined as:
  %Perplexity is algebraically equivalent to the inverse of the geometric mean per-word likelihood and monotonically decreases as the likelihood of the document increases.
%   \vspace{-0.6ex}
  \begin{equation}
 Perplexity(T)=\exp \Bigg\{{- \frac{\sum_{t=1}^{M} \log _{} p(w_t)} {\sum_{t=1}^{M} N_t}} \Bigg\}
 \end{equation}
 Where $T$ is the set of test tweets that are held from the tweet set for building the LDA model; $M$ is size of $T$; $N_t$ is the number of words in a tweet $t$ from tweet set $T$; and $p(w_t)$ is the probability of word distribution in the tweet.
 %A lower value of perplexity indicates better generalization performance though generally speaking, a larger number of topics have a lower perplexity. However, as reported in~\cite{zhao2015heuristic} and \cite{chang2009reading} an LDA model with a lower perplexity can be less semantically interpretable. 
 \cite{zhao2015heuristic}~highlights a few issues with using perplexity to find the appropriate number of topics and proposes additional metric called \textit{rate of the perplexity change (RPC)} for this purpose. Formally, RPC is defined as:
 
  \begin{equation}
 RPC(i)= \left| \frac{P_i -P_{i-1}}{t_i - t_{i-1}}\right |
 \end{equation}
 Where $t_i$ is the number of topics from an increasing sequence of candidate numbers and $P_i$ is the corresponding perplexity. We varied the number of topics from 10 to 150 and observed that RPC is maximized at 70 topics. Thus, we trained the LDA model with 70 topics on a subset of 5M tweets collected from user profiles. For more details see Appendix~\ref{LDA-suplementary}.

 \subsubsection{\textbf{Document embeddings: }}
In order to represent the variable length tweets of neighborhoods with a fixed-length feature vector, we used Doc2vec~\cite{le2014distributed}, an unsupervised framework that learns continuous distributed vector representations for pieces of texts. Details on training the doc2vec model can be found in Appendix~\ref{doc2vec-suplementary}. 
%We tokenized, lemmatized, and removed the stop words  of 5M tweets collected from user profiles. Subsequently, we fit a Doc2vec model on this corpus. We set the vector size to 50. For each neighborhood we concatenate all of the associated tweets and generate the embedding using the trained model. 

\subsubsection{\textbf{Sentiment Distribution:}} 
 %Sentiment analysis or opinion mining - a well studied problem~\cite{vinodhini2012sentiment,medhat2014sentiment,liu2012survey} - is the computational study of people's emotions and opinions toward entities.  
 Sentiment analysis, has been used by researchers for quantifying public moods in the context of unstructured short messages in online social networks~\cite{bertrand2013sentiment}. We also characterize the neighborhoods in our study using the mood of the tweets mentioning a venue located inside the neighborhood. As reported in~\cite{ribeiro2016sentibench} the existing methods for sentiment analysis vary widely regarding their agreement; meaning that depending on the choice of sentiment analysis tool, same content could be interpreted very differently. Thus, we use a combination of several methods to make our framework more robust to the limitations of each method. We used five of the best methods for sentiment analysis~\cite{ribeiro2016sentibench} including Vader~\cite{gilbert2014vader}, Umigon~\cite{levallois2013umigon}, 
SentiStrength~\cite{thelwall2010sentiment}, Opinion Lexicon~\cite{hu2004mining}, and  Sentiment140~\cite{go2009twitter}.  
 %mentioned in Table \ref{tab:senti}.
 We applied the methods on each tweet and normalized the values. Next, we categorized the observed sentiment values in 4 bins and reported the distribution of tweets sentiment for each neighborhood. For more details see Appendix~\ref{sent_supplementary}.

\subsubsection{\textbf{Spatial Distance:}} In order to represent the spatial relationship of the neighborhoods, we computed the geodesic distance between the center points of each pair of neighborhoods. We then normalize the distance values using min-max normalization. 

\subsubsection{\textbf{Common Users:}}
Frequent interaction and the resulting public familiarity engender place-based trust among residents and ultimately the expectation of response to deviant behaviour~\cite{jane1961death}. For a pair of neighborhoods we assume that the greater the number of users that tweeted about both neighborhoods, the higher is the level of the public familiarity of the residents and the more similar are the neighborhoods in terms of level of collective efficacy. Thus, for each pair of neighborhoods we computed the number of users that tweeted about both of the neighborhoods. Then we divided this value by the total number of users that tweeted about at least one of the neighborhoods in the pair. 

 \subsection{Model}\label{model} 
 In this section we discuss our ranking task and model architecture. We use a pairwise approach to automatically generate a ranking of neighborhoods with respect to their collective efficacy. Ranking the objects with a function is equivalent to projecting the objects into a vector and sorting the objects according to the projections. The goal here is to use the extracted features for generating a permutation which is close to the ranking of neighborhoods if sorted by collective efficacy values. In the pairwise approach, the ranking  task is transformed into a pairwise classification problem. In our case, given representations of a pair of neighborhoods $<n_a,n_b>$ the goals is to predict if $n_a$ should be ranked higher than $n_b$ or $n_a$ should come later in the ranking. In the first case, a value +1 is the label to be predicted and in the latter case the value -1 is assigned as the true label. We consider a label value of 0 for a pair of tied neighborhoods since we do not want to move one of them higher or lower in the list with respect to the other one. 
 
We then use this local ordering to generate a global ordering of the neighborhoods. We employed different classifiers for the local ordering task including a neural ranker which is a feed-forward neural network. % performing classification with one task-specific output node. Our neural network has an input layer, three hidden layers, and 100 hidden units at each layer. We use cross entropy loss and Adam algorithm~\cite{kingma2014adam} for optimization. 
%The objective function is to minimize the cross entropy loss of the predicted difference value. We use 
Extensive experiments were conducted to evaluate the effect of the model architecture as well as the predictive power of the features. Our experimental setup and the results are provided in  Sections~\ref{s:EXPERIMENTS} and \ref{s:Results}.
 
  \subsection{Ordering}\label{s:specific setup}
  We train the local ranker model for each pair of the neighborhoods $<n_i,n_j>$ and their corresponding local rank label $r_{ij}$ which can take -1, +1, or 0. For each pair, we also include another training instance $<n_j,n_i>$ as the input and $-r_{ij}$ as the ground truth value. Given the set of tweets associated with neighborhoods in our study we rank the neighborhoods as follows: for every pair $<n_i,n_j>$ we first extract features from tweets of neighborhood $n_i$ and neighborhood $n_j$ then we compute the pairwise features including the spatial distance, and normalized common users count for each pair. We concatenate all the 
  features for every pair mentioned in section~\ref{s:features}. The model then predicts the local ranking for each pair of neighborhoods using the feature representation of each pair.  Let $R(n_i,n_j)$ be the local rank value of neighborhoods $n_i$ and $n_j$ predicted by our model. In order to get the global rank of the neighborhoods, we compute the final score $C(n_i)$ for all neighborhoods by computing: 
%   \vspace{-0.2cm}
  \[
C(n_i) = \sum_{n_i \neq n_j \in N}^{}R(n_i,n_j)
\]

Then we rank the neighborhoods in decreasing order of these scores. The lower the score the lower the degree of collective efficacy of the neighborhood. Similar ranking setup has been used in~\cite{glavavs2015simplifying,paetzold2017lexical,maddela2018word} for substitution ranking.  
 
  \subsection{Evaluation}
  \label{eval}

  We evaluated the accuracy of our model by measuring the agreement between the generated ranking and the ground truth ranking. As mentioned in section~\ref{problem_formulation}, the ranked list of neighborhoods has non-transitive ties. The quality of predicted ranking in this setting can be computed using $\tau_x$ rank correlation coefficient~\cite{emond2002new}. $\tau_b$ is another metric that is used for measuring ranking consensus, however~\cite{emond2002new} uncovers fundamental issues with the usage of $\tau_b$ metric in the presence of ties.
  
  %Agreement between rankings is measured using two main approaches. The first approach is using ranking correlation coefficient, where rankings in full agreement are assigned a value of +1 while rankings in full disagreements are assigned a value of -1. Any ranking which lies in between is assigned a correlation coefficient between these values. The second approach is using a distance metric. In this approach, the distance between any two elements in the set of weak orderings of $n$ objects will be computed. In ~\cite{kemeny1972mathematical}, the authors proposed a set of consistent axioms that a distance measure between two pairs should satisfy. Subsequently, they proved that there is a unique distance function - widely known as Kemeny-Snell distance metric - satisfying all axioms. Further,~\cite{emond2002new} proposed a new ranking correlation coefficient between weak orderings which is equivalent to Kemeny-Snell distance metric. They also proved that Kendall's $\tau_b$ has several issues as a measure of ranking agreement when ties are allowed.  We will define this measure and use it for the evaluation of our framework.
   \subsubsection{The $\tau_x$  rank correlation coefficient}
   \label{tau}
 Let $A$ be a ranking of $n$ objects.  Then~\cite{emond2002new} defines a \textit{weak ordering} $A$ of $n$ objects using the $n \times n$ score matrix. Element ${a_{ij}}$ of this matrix is defined as follows: 

 %~\cite{kendall1948rank} propose a convenient representation for the ranking problem. For any given ranking $A$, of $n$ objects the author defined a $n\times n$ score matrix. Later,~\cite{emond2002new} modified this representation to account for ties in a more mathematically tractable fashion. In a similar way, we define a weak ordering $A$ of $n$ objects using the $n \times n$ score matrix ${a_{ij}}$ as follows: 

\[
  a_{ij} =
  \begin{cases}
                                1& \parbox[t]{.33\textwidth}{if object $i$ is ranked ahead of  or tied with object $j$} \\
                                   -1 & \text{if object $i$ is ranked behind object $j$} \\
  0 & \text{if $i=j$}
  \end{cases}
\]

The $\tau_x$ rank correlation coefficient between two weak orderings $A$ and $B$ is computed by the dot product of their score matrices. 
\[
\tau_x(A,B) = \frac{\sum_{i=1}^{n} \sum_{j=1}^{n} a_{ij}b_{ij} }{n(n-1)}
\]
We further evaluate our proposed framework by computing the $\tau_x $ ranking correlation between the generated ranking and the ground truth ranking. It is important to note that the cumulated gain-based metrics~\cite{jarvelin2002cumulated} such as \textit{Discounted Cumulated Gain (DCG}) and the normalized version of it~(NDCG) widely used in information retrieval literature for examining the retrieval results are not appropriate to evaluate our framework. The main reason being these measures penalize the ranking mistakes more on the higher end of the ranking while devaluing late retrieved items. However, such an objective does not work for our context -  mistakes in ranking on the higher end should be penalized the same as the mistakes in the middle or end of of the list. Thus, employing a measure of ranking agreement is a more appropriate way to evaluate our model. 

%% file: expriments.tex
In this section, we empirically evaluate our hypothesis that ``one can leverage social media data to quantify collective efficacy of neighborhoods".

\input{table_neigh_stats.tex}
\subsection{Dataset and empirical setup}
\label{s:datainfo}
We lever the survey data obtained from the AHDC study. The details about the AHDC study and the computation of collective efficacy for each neighborhood is shared in section~\ref{s:DataCollection}. %In the experiments, unless stated otherwise we exclude those neighborhoods which were reported by less than 5 individuals in the AHDC study. Also, we exclude those neighborhoods which we do not have any tweets associated with a venue located withing their boundaries.
We sorted the neighborhoods based on the number of tweets collected for each of them and (if not stated otherwise) used the top 40\% of neighborhoods in this list in our experiments. This list contains 157 neighborhoods that were mentioned in 3,047 tweets on average. The information on the count of block groups in each set of top k\% of this list as well as minimum, maximum, mean, and median number of tweets collected for each set is reported in Table ~\ref{bgtweetcount}. For more information on the distribution of collective efficacy in each group of the neighborhoods see the Appendix~\ref{collective-efficacy-distribution-suplementary}. We learn our proposed learning to rank model based on 90/10 train/test split of collected tweets. The train/test split also maintains the temporal order where train split is treated as current tweets while test split is treated as future tweets. 
% \vspace{-0.5ex}
\subsection{Baselines for the ranking task}
\label{baselines}
Following are the baselines for the ranking task:
\begin{itemize}[noitemsep,topsep=0pt]
     \item \textbf{Venue count}: Neighborhoods were sorted by the number of venues located in them that were mentioned in the tweets.
    \item \textbf{Population}: Neighborhoods were sorted by total population of them. The values are extracted from the 2013 report of the United States census bureau\footnote{https://www.census.gov}. 
     \item \textbf{Tweet count}: We sort the neighborhoods by number of tweets that mentioned a venue located in them.
     \item \textbf{User count}: We sort the neighborhoods by number of users that tweeted about a venue located in them.
    \item \textbf{Random}: 
    %We randomly permute the list of neighborhoods and compute $\tau_x$ ranking correlation coefficient.
    We generate 100 random permutations of the neighborhoods and report the average $\tau_x$.
    %  \item \textbf{Coordinates}: We sort the neighborhoods by the coordinates of the center of their geo polygon shapes. This means that we sort the neighborhoods by their longitude value and for two neighborhoods with same longitude with sort them by their latitude value. We repeat this process reversing the order of sorting based on latitude and longitude. We report the average $\tau_x$ of the two permutations.
\end{itemize}

\subsection{The classifier for the local ranking task}
\label{clssifiers-param-discription}
Since we rely on pairwise learning to rank, we experiment with below classifiers for our local ranking task. The parameters are tuned using  grid search with cross-validation parameter set to 5 and scoring function set to `f1'.  
\begin{itemize}[noitemsep,topsep=0pt]
    \item \textbf{Logistic Regression (LR)}: The estimator penalty is set to `L1' and the inverse of regularization strength is set to 0.1. 
    \item \textbf{Support Vector Machine (SVM)}: The kernel is set to `rbf', the penalty parameter C is set to 1, and the gamma kernel coefficient for rbf is set to 0.1. 
    \item \textbf{Random Forest (RF)}: The number of estimators is set to 200. The minimum number of samples required to be at a leaf node is set to 5, and the function to measure the quality of a split is set to `gini'.  
    \item \textbf{Multi-layer Perceptron (MLP)}: We use a feed-forward neural network with 3 hidden layers and 100 units at each hidden layer, and a task-specific output layer. We use cross entropy loss and Adam algorithm~\cite{kingma2014adam} for optimization. 
\end{itemize}

As discussed in section~\ref{problem_formulation} we define the tied neighborhoods as the ones having a significantly small difference in collective efficacy value. Tied neighborhoods are considered interchangeable in the ranking. We define the ties based on a threshold on collective efficacy difference. We compute the standard deviation of the collective efficacy value of the neighborhoods in our study and define our threshold based on different coefficients of the standard deviation of the collective efficacy. We vary the coefficient from 0 to 1 with 0.2 increments and evaluate the ranking consensus using a ranking correlation metric discussed in section~\ref{eval}. More details on tied neighborhoods in shared in Appendix~\ref{ties-supplementary}. The results are discussed in section~\ref{s:Results}.

%% file: table_neigh_stats.tex
\begin{table}[]

\resizebox{\linewidth}{!}{

\begin{tabular}{c||c|c|c|c|c}
\textbf{\% of top tweeted} & \textbf{Neighborhood } & \textbf{Min. }  & \textbf{Median} & \textbf{Mean} & \textbf{Standard Deviation} \\
 \textbf{ Neighborhoods} & \textbf{Count} &  & & &\textbf{of Collective Efficacy} \\

\hline

\textbf{20} & 78 & 490  & 1394.5 & 5903.5 & 0.1826 \\
\textbf{40} & 157 & 110  & 473 & 3047.7  & 0.1821\\ 
\textbf{60} & 235 & 39  & 197 & 2058.8 & 0.1962 \\ 
\textbf{80} & 314 & 14  & 110 & 1546.9 & 0.2038\\
\textbf{100} & 393 & 1  & 63  & 1237.2 & 0.2063 \\ 

\end{tabular}
}

\caption{The neighborhood count, minimum, mean, and median number of tweets at each set of top k\% neighborhoods of the sorted list of the block groups based on their tweet count. The maximum number of tweets in each set of top k\% neighborhoods is 98,951. \label{bgtweetcount}}
\label{t:set_info}
\end{table}

\begin{comment}

\hline
\multicolumn{1}{|l|}{\textbf{Percentage of top tweeted Neighborhoods}} & \multicolumn{1}{l|}{\textbf{Neighborhood Count}} & \multicolumn{1}{l|}{\textbf{Min}} & \multicolumn{1}{l|}{\textbf{Max}} & \multicolumn{1}{l|}{\textbf{Median}} & \multicolumn{1}{l|}{\textbf{Mean}} \\ \hline
\end{comment}

%% file: results.tex
\begin{table*}[]
\centering
\resizebox{0.98\linewidth}{!}{
\begin{tabular}{l||l|c|c|c|c|c|c|cc}

{\textbf{ID}}  & \multicolumn{1}{l|}{\textbf{Models}}                               & \textbf{0} & \textbf{0.2} & \textbf{0.4} & \textbf{0.6} & \textbf{0.8} & \textbf{1}  & \textbf{AUC-ERC} \\ \hline
1 & \textbf{Random}                                                         & 0     &  0.09       & 0.20       & 0.30       & 0.40      & 0.49      & 0.25      \\ \hline

2 & \textbf{Coordinates}                                                         & -0.04     &  0.05       & 0.17       & 0.27       & 0.37      &  0.47   & 0.22       \\ \hline 

3 & \textbf{User count}                                                         & 0.03     &  0.13       & 0.25       & 0.35      & 0.45      & 0.53      & 0.292     \\ \hline

4 & \textbf{Tweet count}                                                         & 0.04     &  0.14       & 0.25       & 0.35      & 0.45      & 0.53      & 0.295      \\ \hline

5 & \textbf{Population}                                                         & 0.05     &  0.15       & 0.26       & 0.36      & 0.46      & 0.55      & 0.306     \\ \hline

6 & \textbf{Venue count}                                                         & 0.09     &  0.19       & 0.3       & 0.4      & 0.49     & 0.58      & 0.342     \\ \hline
7 & \textbf{Doc2vec + Sentiment}                                             & 0.3539     & 0.4504       & 0.5256       & 0.6347       & 0.713        & 0.7635   & 0.5764       \\ \hline
8 & \textbf{Doc2vec + Sentiment + Common Users}                              & 0.3698     & \textbf{0.461}        & \textbf{0.5497}       & 0.6043       & 0.7033       & 0.7642     & 0.5770       \\ \hline
9 & \textbf{Doc2vec + Sentiment + Common Users + Topics}                     & 0.368      & 0.4367       & 0.5425       & \textbf{0.6412}       & 0.6988       & 0.7647    & 0.5771       \\ \hline
10 & \textbf{Doc2vec + Sentiment + Common Users + Topics  + Distance}         & \textbf{0.3748}     & 0.4565       & 0.5388       & 0.6322       &\textbf{ 0.7207 }      &\textbf{ 0.7735}    & \textbf{0.5844}       \\ \hline
11 & \textbf{Doc2vec + Sentiment + Common Users + Topics  + Distance + Tfidf} & 0.3686     & 0.4597       & 0.5207       & 0.6294       & 0.6957       & 0.7666     & 0.5746      
\end{tabular}
 }
\caption{The drilldown of our proposed model. The comparison of Kendall $\tau_x$  and area under the collective efficacy ranking curve(AUC-ERC) of the baselines and our proposed framework. Top 40\% of highly tweeted block groups were included in this experiment. Random forest was used for local ordering task. %Doc2vec  , the densest content features was used in all combinations and other features were added one at the time. due to its consistent 
%The AUC-ERC is maximized when doc2vec, topic distribution, normalized common users count, sentiment, and spatial distance were used to characterize the neighborhoods.
\label{t:ablation_study}}
% \vspace{-3.0ex}
\end{table*}	
\begin{figure}[h]
\centering
 \includegraphics[width=1\linewidth,height=6.4cm]{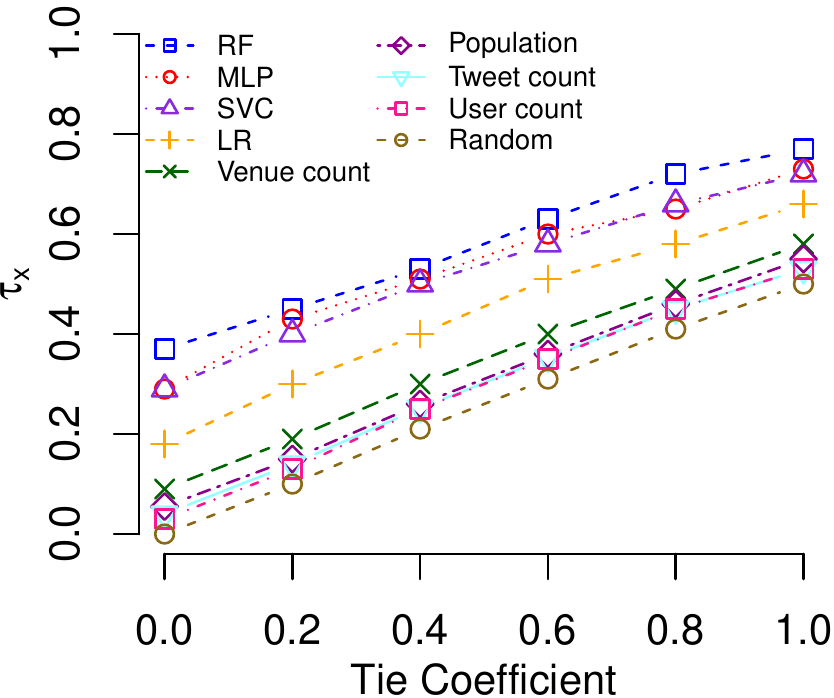}
\caption{Ranking performance of our proposed model and the baselines. We used 4 classifiers for local ranking module. The x axis indicates the coefficient that is multiplied by standard deviation to make the tie threshold. The standard deviation of the ground truth collective efficacy for the 157 block groups included in this experiment is 0.18.}
\label{fig:rank_perf}
\end{figure}
\subsection{Ranking performance}
In this section, we present the results of ranking agreement of the permutation generated by our framework using 4 different classifiers when the most informative combination of features discussed in Section~\ref{s:features} were used. More specifically, we used doc2vec, distribution of topics, distribution of sentiment, normalized common user count, and spatial distance to characterize each pair of neighborhoods in our study. More details on parameter setting for classifiers and feature analysis is presented in Section~\ref{clssifiers-param-discription} and Section~\ref{s:model_drill_down}.
% The ranking performance of the baselines described in Section~\ref{baselines} is also depicted. 
% We used 4 different classifiers including random forest, logistic regression, support vector classifier, and multi-layer perceptron for our local ranking task.  
As shown in the Figure~\ref{fig:rank_perf}, our framework even when used with a linear classifier such as logistic regression outperforms the baselines by at least 20\%. Also, it can be seen that random forest closely followed by multi-layer perceptron is consistently giving better ranking correlation results in comparison to other classifiers.  

\subsection{Model drill down}
\label{s:model_drill_down}

In this section we discuss our experiments related to the effect of each feature discussed in Section~\ref{s:features} on ranking. 
\begin{comment}
%We noticed that a feature combination is not consistently doing better or worse for all tie coefficients. To account for this, we measured the area on collective efficacy ranking correlation curve(AUC-ERC) for each of the factor combinations.  The results are presented in Table~\ref{t:ablation_study}. 
%We proposed three content features, namely doc2vec, distribution of topics, and TF-IDF of crime lexicon. Among these content features doc2vec is the densest. Thus, we used doc2vec as the baseline and kept adding other features one at a time. 

% \begin{figure}[t]
% \resizebox{0.9\linewidth}{!}{
%     \centering
%     \includegraphics{plots/box.pdf}
%     }
%     \caption{The box plot of $\tau_x$ values for TF-IDF,  distribution of topics and doc2vec baselines on Top 40\% of tweeted neighborhoods with tie threshold from 0 to 1 with an interval of 0.2. Three classifiers including random forest, multi layer perceptron, and logistic regression were used to conduct overall of 72 experiments per content feature. }
%     \label{fig:box_plot}
% \end{figure}

%From Table ~\ref{t:ablation_study}, we see that all the baselines perform poorly based on AUC-ERC score. 
\end{comment}
To determine the best context feature, we experimented with features in this group namely TF-IDF of crime lexicon, topic distribution of urban activities, and doc2vec on the top 40\% highly tweeted neighborhoods. We performed experiments with all different combinations of our 3 content factors. %resulting in a full $2^3-1$ factorial design experiment. 
Each content feature is enabled in 3 combinations and disabled in 3 other corresponding paired combinations.
Each factorial experiment was conducted using 3 classifiers for the local ranking module including random forest, multi layer perceptron, and logistic regression. We repeated this process for 6 tie coefficients. Tie coefficients varied from 0 to 1 with an interval of 0.2. This resulted in 54 experiments in which a content feature is enabled and 54 experiments in which a content feature is disabled. %The box plot of the ranking correlation of these 72 experiments is depicted for each feature. 
We observed that adding doc2vec increases the ranking performance and this boost is statistically significant~(Wilcoxon signed-rank test with p-value~\textless~ 0.001)~\cite{wilcoxon1970critical}. However, this was not the case for two other content features. 
%more than adding TF-IDF and topics and the difference is statistically significant~(Mann-Whitney U Test~\cite{mann1947test}) with a p-value of 0.009 and 0.004 respectively.
%as the mean of $\tau_x$ for doc2vec is above the mean of $\tau_x$ for other features. 
The box plot of these experiments is shared in Appendix~\ref{results-suplementary}. Next, we examine the impact of additional features along with doc2vec on the 
%Next, we evaluate each feature individually and evaluate their ranking performance. We see that Doc2vec consistently outperform other single features in the ranking performance with  AUC-ERC of 0.57. We also see that the single dimensional feature Distance perform poorly in ranking task when random forest classifier is used. However, we also see that when Distance feature is added in baseline number 11 in Table ~\ref{t:ablation_study}, resulting in baseline number 12, the performance of resultant model increases as measured by AUC-ERC score. This suggests that Distance feature when considered alone is not highly discriminative however when augmented with other features provides a different signal not present in other features.
%Since, Doc2vec feature achieves the best AUC-ERC score, we examine the impact of additional features on the 
ranking performance. In each experiment we computed the ranking correlation of the generated ranking with the ground truth ranking with tie coefficient values ranging from 0 to 1 with interval of 0.2. The results are presented in Table~\ref{t:ablation_study}. As it can be seen from the Table, regardless of choice of feature combination or tie coefficient our model consistently outperforms the baselines. 
For summarizing the ranking correlation results, we rely on \textbf{AUC-ERC}. AUC-ERC is area under the curve of graph created by plotting tie coefficients against $\tau_x$. 
From Table~\ref{t:ablation_study}, we see that models 7 to 11, show better AUC-ERC score than the baselines. Model number 10 achieves the highest AUC-ERC score. We see that model 11 which includes all the features is not the best performing model. We conjecture that this behaviour is due to the over-fitting of the model on the training set. The TF-IDF feature has a dimensionality of 100 and the classifier might learn a function to predict the local rank between pair of neighborhoods based on few crime lexicon words (e.g. gun, shooting) in the training set. However, the test set might not contain those words on which classifier learned the function thereby resulting in wrong prediction. To summarize, using Model 10 as our proposed model, our generated ranking of neighborhoods achieves 0.77 Kendall tau-x ranking agreement with the ground truth ranking. Our results are between 20\% to 37\% better than the baselines depending on choice of the tie threshold.

%As shown in the table, combination of sentiment (valence), doc2vec and topic distribution (content), common users (topology), and distance (spatial) capturing different aspects of neighborhoods, performs the best in the ranking task. However, since frequency of mentioning a crime related word is most of the tweets is often close to zero, adding this sparse feature results the ranking consensus to drop. Note that doc2vec, distribution of topics, and frequency of the crime lexicon are all high dimensional features while common users, distance, and sentiment have much lower dimensional. Thus, the value they add to the prediction task is expected to be small. 

\subsection{Effect of data availability}

\begin{figure}[]
 \includegraphics[width=1\linewidth,height=5cm]{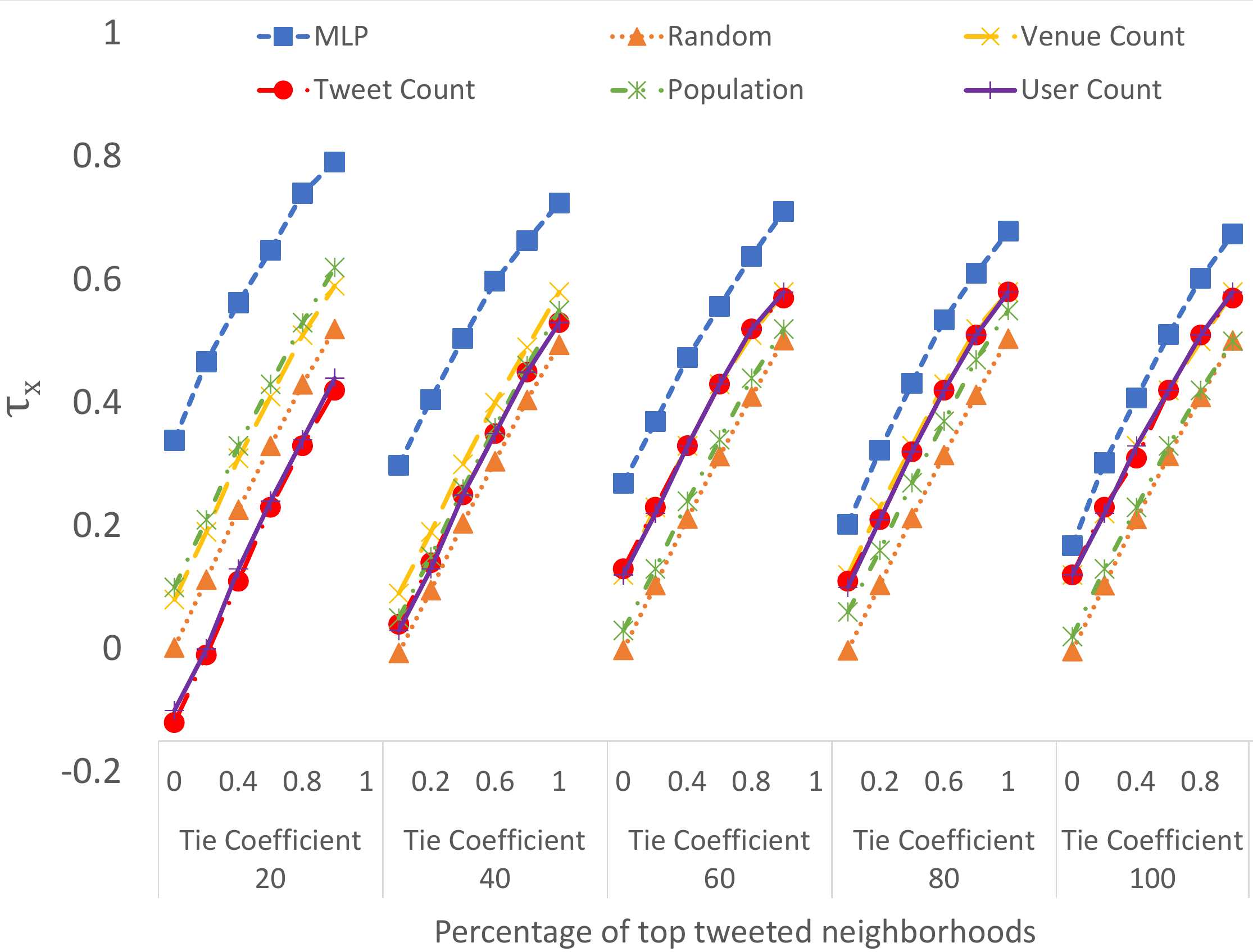}
\centering
\caption{Ranking performance of our framework and the baselines on different sets of neighborhoods. The sets are defined based on the number of collected tweets. The x axis indicates the tie coefficient. Tie threshold is computed by multiplying the standard deviation of collective efficacy by the tie coefficient. The standard deviation of each set is reported in Table~\ref{t:set_info}.} 
\label{fig:topk}
\end{figure}

In this section, we explored to what extent the result of ranking consensus is related to the amount of data we have for each neighborhood. With this in mind, we solved the ranking tasks for different set of block groups. These sets are introduced in Section~\ref{s:datainfo}.
%defined based the number of tweets we have for each neighborhood in our study. For instance, top 20\% indicates the top 20\% of the neighborhoods that were highly tweeted. The neighborhoods in this set had 3047 tweets on average. More details on the sets can be found in Section~\ref{s:datainfo} and Table~\ref{bgtweetcount}.
We used our ranking framework with MLP as the classifier to rank each set. As indicated in Figure~\ref{fig:topk} the more the amount of data we have for the neighborhoods in our study, the higher is the ranking consensus of the generated ranking and the ground truth ranking.

\begin{comment}
3 & \textbf{common users}                                                         & 0.06271     &  0.1657       & 0.1786      & 0.2792       & 0.3785      &  0.4964   & 0.2563     \\ \hline
4 & \textbf{distance}                                                         & -0.1084     &  0.04507      & 0.1502      & 0.3865       & 0.3901      &  0.5151    & 0.2350      \\ \hline
5 & \textbf{sentiment}                                                         & 0.0796     &  0.1905      & 0.2897      & 0.3908       & 0.4878      &  0.5753    & 0.3373      \\ \hline

6 & \textbf{tf-idf}                                                         & 0.1286     &  0.2745      & 0.3539      & 0.4390       & 0.5082      &  0.6049  & 0.3885      \\ \hline

7 & \textbf{topics}                                                         & 0.1407     &  0.2395      & 0.3392      & 0.4274       & 0.5218      &  0.6266    & 0.3823      \\ \hline
8 & \textbf{doc2vec}                                                         & 0.3505     & 0.4486       & 0.5283       & 0.6165       & 0.7033       & 0.7601    & 0.5704       \\ \hline \hline \hline 
\end{comment}

%% file: conclusion.tex
In this paper, we focused on the problem of costly computation of collective efficacy values for the neighborhoods. With the help of extensive experiments, we showed that this problem can be addressed by leveraging the social media data. %As the neighbourhoods changes with time, the collective efficacy of those neighbourhoods also changes.
Our proposed framework allows frequent and less costly access to collective efficacy values of the neighborhoods. In the future, we plan to leverage data from other sources (e.g., additional social forums and census) to improve our model. Additionally, we plan to explore the ego-net of users on social media and weigh high importance to tweets of users who are more familiar with a particular neighbourhood. Our proposed framework can act as an early warning system to capture the transformations in the neighborhoods' composition. This potentially can assist regulators and policymakers to prioritize resources, monitor neighborhood safety, and upkeep. It is our intent to release the tweet IDs as well as the ground truth collective efficacy values of the neighborhoods once this work is published.

\begin{comment}
    proposed a framework that given the tweets associated with the neighborhoods can rank them based on the collective efficacy value of the neighborhoods. We characterized each neighborhood in our study area using features extracted from social media. These features are related to the type of urban activity, visible signs of crime and violence, the familiarity of residents with one another, and the public mood of the neighborhood. Using our proposed feature, we solved a pairwise ranking task and  generated a permutation of neighborhoods that has a Kendall $\tau_x$ of 0.77 ranking agreement with the ground truth collective efficacy rank achieved from a survey study. 
\end{comment}

%% file: ack.tex
This material is based upon work supported by the National Institute of Health (NIH) under Grant No NIH-1R01 HD088545-01A1. Any opinions, findings, and conclusions in this material are those of the author(s) and may not reflect the views of the respective funding agency.

%% file: Suplementary_material.tex
\subsection{Data Collection}
\label{datacollection-suplementary}
We collected a significant amount of data from Twitter. We used Snowball sampling to identify Twitter accounts of local citizens. Crawling publicly available tweets from user profiles enables us to collect significantly more amount of data in comparison to collecting streaming real-time tweets of the Columbus area.  
For the purpose of our study, 63 Twitter accounts that mostly posted news and information about Columbus city were identified and used as the seed users. Many local residents follow such accounts to stay informed about the local events~\cite{kwak2010twitter}. The seed set included the twitter account of several organizations including major universities, recreational centers, medical centers, newspapers, local bloggers, local reporters, police, libraries, restaurants as well as the local sports teams. Using Twitter's streaming API, the followers of the seed accounts were collected. Following this step, we explored user's profiles and identified 54K public profiles that marked their locations as Columbus or one of the suburban areas included in the AHDC study. The AHDC study area included several populous suburbs.  
%\textit{Whitehall},\textit{ Bexley}, \textit{Upper Arlington}, \textit{Grandview Heigths}, \textit{Worthington}, \textit{Hilliard}, and \textit{Dublin}.  
Collectively, 50 million publicly available tweets were collected from these accounts. In another wave of data collection, we collected publicly available geo-tagged tweets for a period of May-August of 2018. This resulted to additional 2.8 million tweets. Next, LNEx was used for location name extraction from tweets and associating tweets to neighborhoods. 
There are cases in which ambiguous locations were reported by LNEx. In our study, we exclude tweets containing ambiguous location entities. 
The location ambiguities were observed in a following cases: 
\begin{itemize}[noitemsep,topsep=0pt]
    \item A location entity may have several matches in the gazetteer. For example, \textit{Holiday Inn} and \textit{Gamestop}. 
    \item A location entity having a single gazetteer entry can potentially refer to a huge area. For example, gazetteer entry of Trans-Siberian Highway in Russia spans from St. Petersberg to Vladivastok. Such entities cannot be mapped to a single neighborhood. 
    \item Location entities extracted by LNEx having a gazetteer entry but not referring to a location in the context. For example, \textit{American Girl}, \textit{Modern Male} etc. 
\end{itemize}

Such mentions were identified manually and excluded from the study. This pruning step resulted in 4846 unique locations in the area that were spotted in 545k tweets and were mapped to 424 neighborhoods.

\begin{figure}[t]
 \includegraphics[width=0.9\linewidth,height=5.5cm]{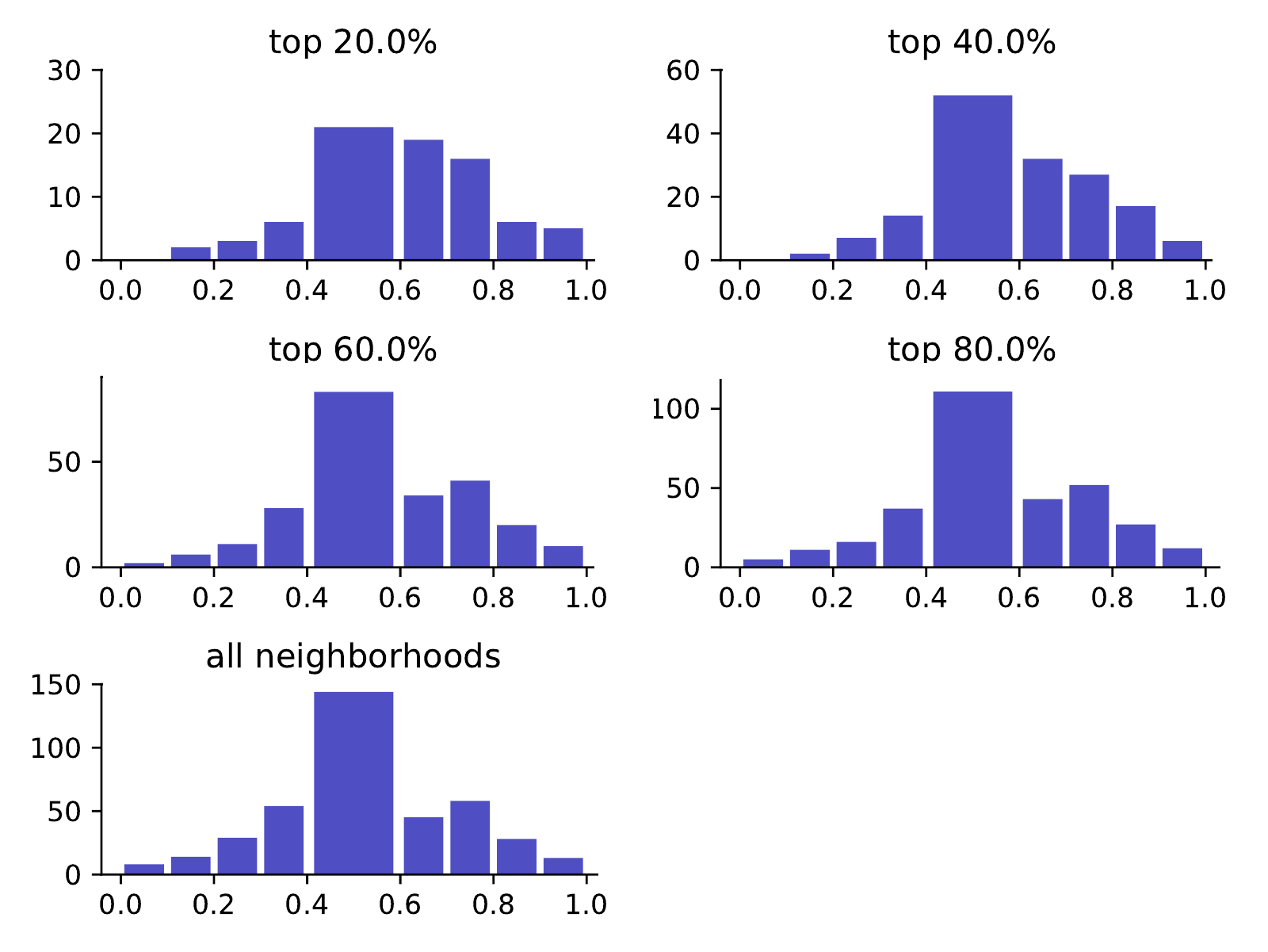}
\centering
\caption{Distribution of the collective efficacy values in each set of neighborhoods.  The collective efficacy values are computed from the survey study and are normalized in the 0 to 1 range. Refer to Section~\ref{s:DataCollection} for more details.
}
\label{fig:distribution}
\end{figure}

\subsection{TF-IDF of crime related words}
\label{tfidf-supplementary}
We tokenized each tweet in our train set preserving the hashtags, handles, and emojis as separate words. We then removed the stopwords and lemmatized the tokens. Bigrams of the tweets were added to the token set. The top 100 crime-related terms that had the most frequency across the tweets were chosen as our vocabulary set. For the test set, we concatenated all the tweets in each neighborhood to get a single corpus per each neighborhood. We then, transformed each corpus to get the corresponding term-document vector.  

\subsection{Distribution of spatio-temporal urban activities}
\label{LDA-suplementary}
\begin{figure}[t]
\resizebox{1\linewidth}{!}{
\includegraphics[]{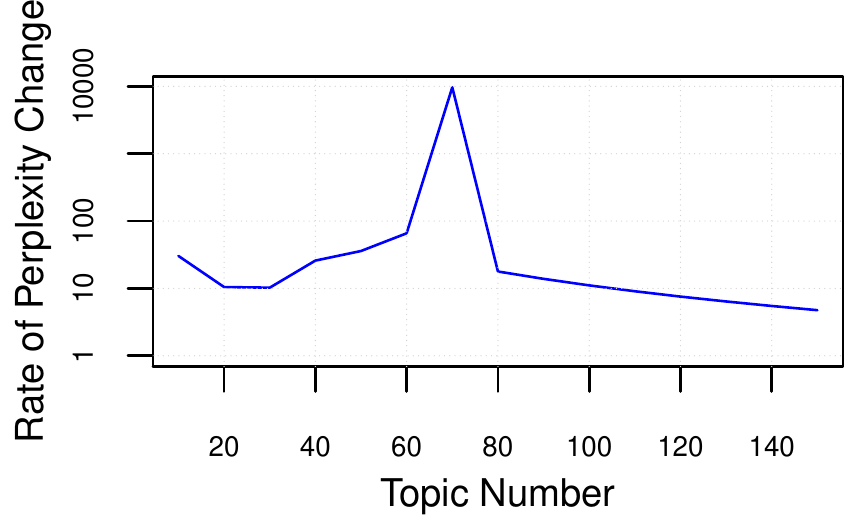}}
\centering
\caption{RPC was used to determine the appropriate number of topics. RPC is maximized for 70 topics.
\label{fig:rpc}
}
\end{figure}
Prior to feeding the corpus to the LDA module we tokenized the tweets using a tokenizer adapted for tweets\footnote{We used the open source tokenizer presented in https://github.com/erikavaris/tokenizer}, removed stop words, lemmatized the tokens, and added the bi-grams that appeared in more that 20 tweets to our set of tokens. Next, we removed the words that appeared in less than 20 tweets~(rare words) or more than 50\% of the tweets. Employing RPC, we used an increment of 10 and varied the number of topics from 10 to 150 and trained LDA model on a corpus of 5M tweets collected from users' profiles. As depicted in Figure~\ref{fig:rpc} RPC in maximized at 70 topics. Thus, we used 70 as the optimal number of topics for our model. 

\subsection{Document Embedding}
\label{doc2vec-suplementary}
We tokenized, lemmatized, and removed the stop words  of 5M tweets collected from user profiles. Subsequently, we fit a Doc2vec model on this corpus. We set the vector size to 50. For each neighborhood we concatenate all of the associated tweets and generate the embedding using the trained model. 

\subsection{Sentiment Distribution}
\label{sent_supplementary}
We applied the 5 sentiment analysis tools to each tweet and normalized the values in a range of -1 to 1. Most of these tools predict the sentiment value using a predefined lexicon. Thus, they cannot perform accurately in the absence of sentiment lexicons in the tweets. To account for this, for each tweet, we only consider the non-zero outputs and compute the average value of them. Subsequently, we use a binning step to put the tweets associated with a neighborhood in four bins - highly negative, negative, positive, and highly positive. We normalized the value of bins by dividing the counts by the total number of tweets of the neighborhood. At the end of this step, for each neighborhood, we report the distribution of sentiment of all the tweets mentioning a venue located inside the boundaries of the neighborhood.
\subsection{Distribution of Collective Efficacy}
\label{collective-efficacy-distribution-suplementary}
The distribution of collective efficacy has been presented in Figure~\ref{fig:distribution}. As it can be seen in the plot, in all of neighborhood sets, the distribution of collective efficacy ground truth values in approximately similar to set of all neighborhoods. Also, it can be seen that most of the block groups have a collective efficacy value between 0.4 and 0.6.

\begin{figure}[t]
\resizebox{0.9\linewidth}{!}{
    \centering
    \includegraphics{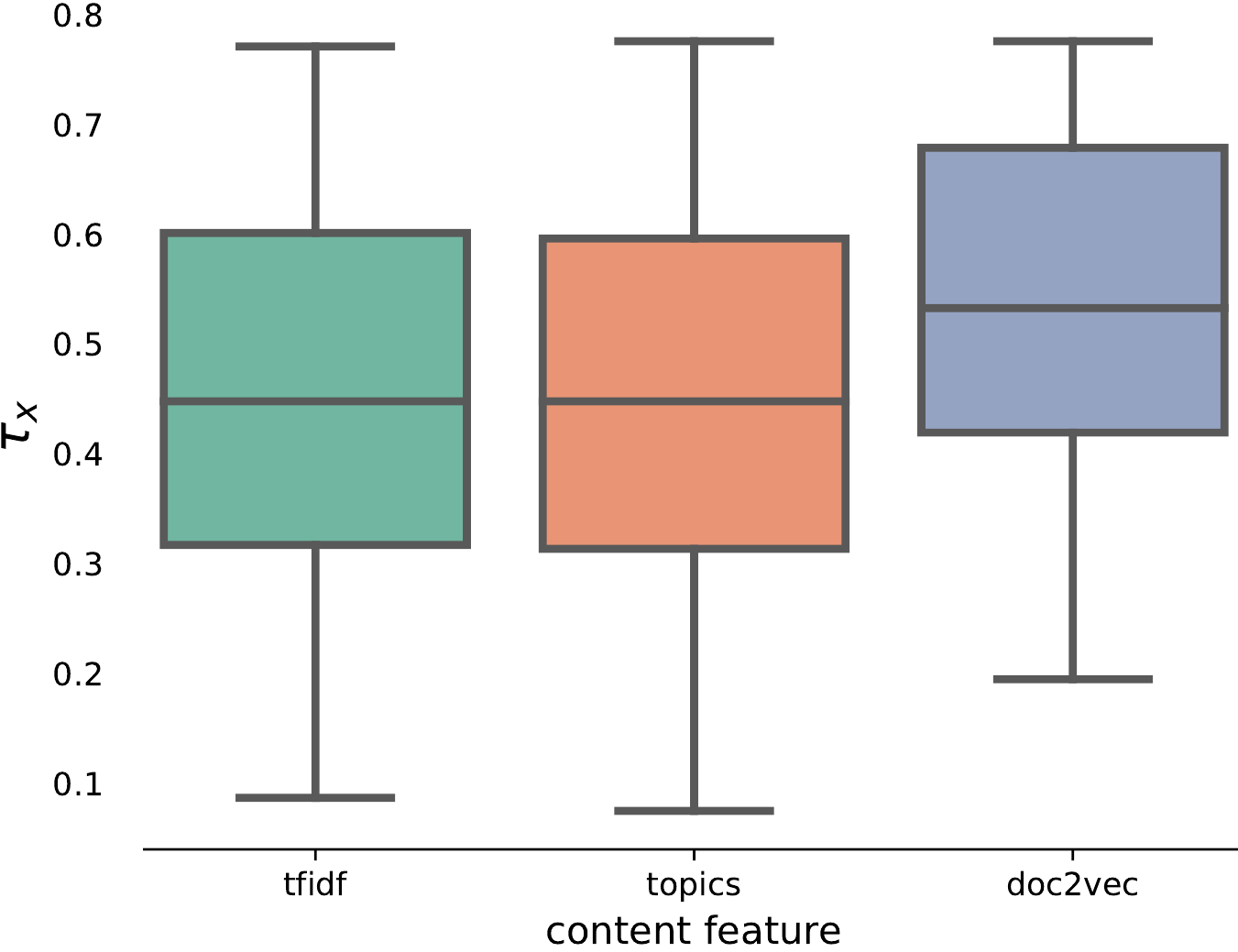}
    }
    \caption{The box plot of $\tau_x$ values for TF-IDF,  distribution of topics and doc2vec baselines on Top 40\% of tweeted neighborhoods with tie threshold from 0 to 1 with an interval of 0.2. Three classifiers including random forest, multi layer perceptron, and logistic regression were used to conduct overall of 54 experiments per content feature. }
    \label{fig:box_plot}
\end{figure}

\subsection{Tied Neighborhoods}
\label{ties-supplementary}
\begin{figure}
 \includegraphics[scale=0.42]{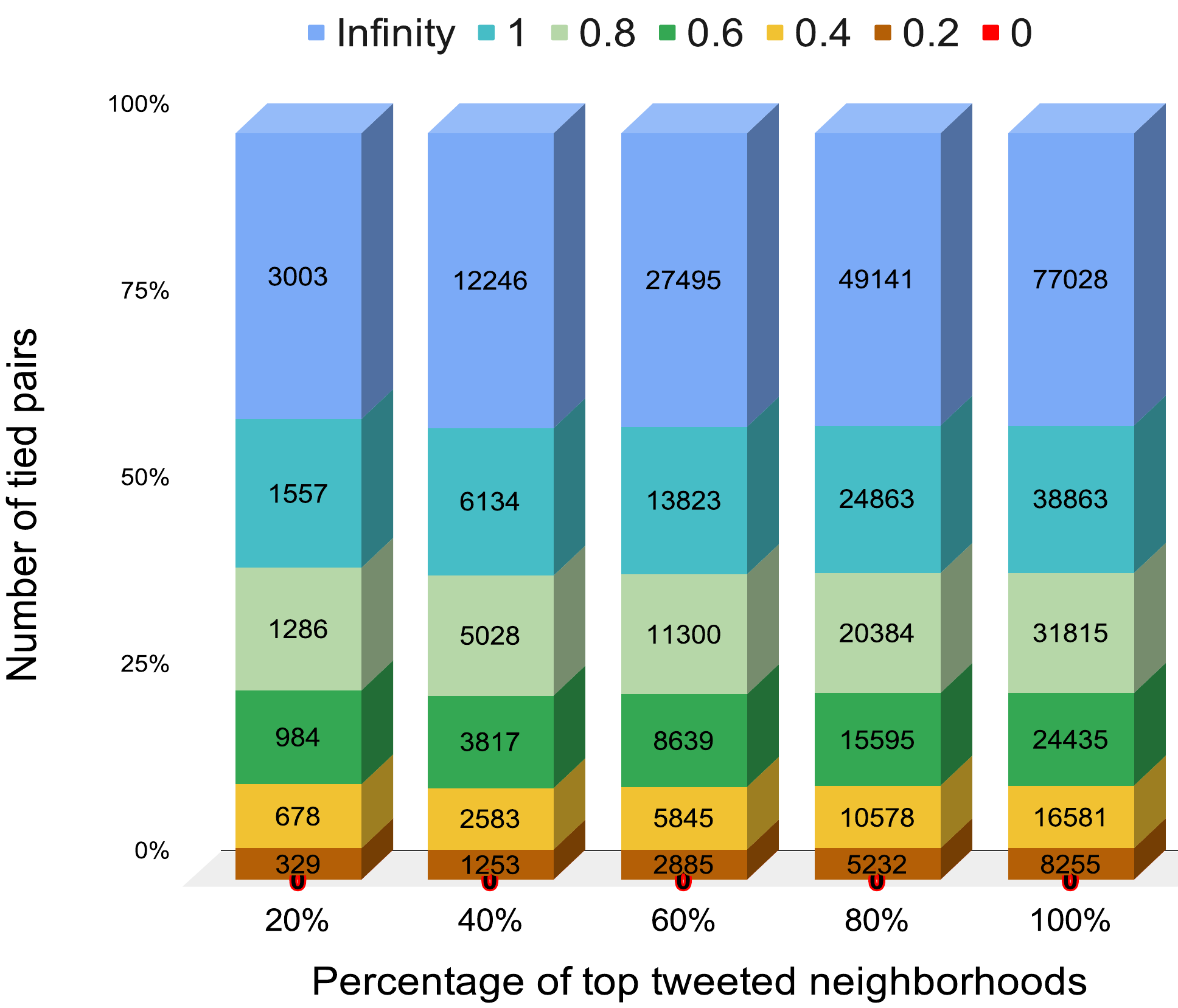}
\centering
\caption{Number of tied neighborhood pairs at each tie coefficient. Tied neighborhoods are not ranked against each other. For N number of neighborhoods at each set, the number of paired is $N \times N-1$. Total number of pairs at each tie threshold is shown with label "infinity". By increasing the tie threshold, the number of tied neighborhoods increases. }
\label{fig:pairs_count}
\end{figure}
As discussed in section~\ref{problem_formulation} we define tied neighborhoods as the ones having a significantly small difference in collective efficacy value. Tied neighborhoods are considered interchangeable in the ranking. We define the ties based on a threshold on collective efficacy difference. We refer to this value as "Tie Threshold". We compute the standard deviation of the collective efficacy value of the neighborhoods in our study and define our threshold based on different coefficients of the standard deviation of the collective efficacy. We refer to these coefficient as "Tie Coefficient". We vary the coefficient from 0 to 1 with 0.2 increments and evaluate the ranking consensus using a ranking correlation metric discussed in section~\ref{eval}. The number of neighborhoods that are considered as "tied" at each tie threshold is shown in Figure~\ref{fig:pairs_count}. As indicated in the plot, by increasing the tie threshold, number of tied pairs increases.

\subsection{Results}
\label{results-suplementary}
In order to find the best context feature, we experimented with features in this group namely TF-IDF of crime lexicon, topic distribution of urban activities, and doc2vec on the top 40\% highly tweeted neighborhoods. We performed experiments with all different combinations of our 3 content factors. %resulting in a full $2^3-1$ factorial design experiment. 
Each content feature is enabled in 3 combinations and disabled in 3 other corresponding paired combinations.
We conducted each factorial experiment with 3 classifiers for the local ranking module including random forest, multi layer perceptron, and logistic regression. We repeated this process for 6 tie coefficients. Tie coefficients varied from 0 to 1 with an interval of 0.2. The number of tied neighborhoods at each tie coefficient is shown in Figure~\ref{fig:pairs_count}. The cross product of these parameters resulted to $3\times 3\times6 = 54$ experiments in which a content feature is enabled. The box plot of the observed ranking consensus for these 54 experiment for each content feature is presented in Figure~\ref{fig:box_plot}. As it can be seen in the figure, by characterizing neighborhood's tweets using doc2vec we consistently generate better rankings in comparison to TF-IDF of crime lexicon and topic distribution of urban activities.